\documentclass[proceedings]{JHEP3}

\PrHEP{PrHEP hep2001}
\conference{International Europhysics Conference on HEP}

\usepackage{epsfig}
\newcommand{\be}{\begin{equation}}
\newcommand{\ee}{\end{equation}}
\newcommand{\bS}{\mathbf{S}}
\newcommand{\bR}{\mathbf{R}}
\newcommand{\bM}{\mathbf{M}}

\title{Asymptotic freedom with discrete spin variables?}

\author{Peter Hasenfratz and 
        Ferenc Niedermayer\thanks{On leave from E\"otv\"os
          University, Budapest} \\
        Institute for Theoretical Physics, University of Bern,\\
        Sidlerstrasse 5, CH-3012 Bern, Switzerland \\
        E-mail: \email{hasenfra@itp.unibe.ch, niederma@itp.unibe.ch}}

\abstract{We study the critical behaviour of the 2d dodecahedron spin
model and investigate the conjecture that the discrete  model
describes the same continuum theory as the O(3) non-linear sigma
model. In particular, we found that the anisotropy
of the magnetization $A(z)$ measured in a fixed physical volume
decreases with increasing correlation length, at least up
to $\xi \approx 1000$.}

\begin{document}


One of the most studied quantum field theories is the 2d non-linear
sigma model. The lattice regularized version for the O(3) case
is given by the action
\be
{\cal A}(\bS) = 
-\frac{1}{g} \sum_{x,\mu} \bS(x) \bS(x+\hat{\mu})
\ee
where $\bS(x)$ is a unit O(3) vector, $\bS^2(x)=1$.
The model is widely believed to be asymptotically free (AF)
and to have a dynamically generated mass gap.

Here we discuss the analogous model when the spin variables are
allowed to take only discrete values pointing towards the vertices 
of a Platonic regular solid, e.g. icosahedron (I) or dodecahedron (D).
Accordingly, the symmetry of the discrete model is reduced to a discrete
subgroup ${\cal D}$ of O(3) (in the actual case, to the icosahedral
group with 60 elements).
At small $g$ (low temperature) the discrete symmetry is spontaneously
broken while at large $g$ the system is in the symmetric phase
under ${\cal D}$. Numerical evidence shows that there is a second order 
phase transition at some $g=g_c > 0$.

A few years ago Patrascioiu and Seiler \cite{PS-98a}
observed that within their statistical errors the physical results
for the dodecahedron model for $g\searrow g_c$ 
are consistent with those of the O(3) model and conjectured that 
the two models are equivalent in the continuum limit.
The authors saw this observation as an argument supporting
their unorthodox view that the O(3) model is not AF.

The numerical evidence of \cite{PS-98a} inspired us to study this question.
In \cite{HN-01a} we pointed out that the equivalence is not in
contradiction with the O(3) model being AF. Of course, for the
discrete model this is not a statement about the {\em bare}
coupling (which goes to a non-zero constant in the continuum limit)
but about some {\em physical}  running coupling,
say the L\"uscher-Weisz-Wolff coupling $\overline{g}(L)$ \cite{LWW}.
In \cite{HN-01a} we measured the renormalized zero-momentum
4-point coupling $g_R$ and have found that the I and D models
approach the O(3) model with increasing $\xi$.
In that work we went up to correlation length\footnote{Here
and below by $\xi$ we denote the infinite volume correlation
length. When not measured directly it is calculated from
the FSS function of \cite{CEPS-95}.}
$\xi\approx 300$ and reached precision of O(0.1\%). 

In \cite{HN-01a} we also compared the finite size scaling (FSS)
function of the LWW coupling in the discrete models 
to that of the O(3) model, with a similar conclusion. 

Recently Caracciolo et al. \cite{CMP-01a} argued that the operator
which has to be added to the O(3) action to break the original 
O(3) symmetry down to ${\cal D}$ is a relevant operator and concluded
that the equivalence cannot hold. They suggested that
the numerical evidence is misleading and the discrete model should
belong to a different universality class, although they expected
this to show up only at $\xi \gtrsim 200$.
At this conference they have presented measurements of the FSS
function and concluded that the discrete model starts to depart
from the O(3), although at much larger correlation length, 
$\xi \approx 10^5$ \cite{CMP-01b}. 
We shall reflect on their arguments in the concluding section.


For the discrete models we have measured $g_R(z)$, the renormalized
coupling at a fixed finite physical volume, $z=L/\xi(L)$,
where $\xi(L)$ is the ``second-moment correlation length''.
Fig.~\ref{gr_dev} shows the deviation of $g_R(z)$ from O(3) at $z=2.32$ 
as a function of $1/\xi(L)$, both for the I and D case \cite{HN-01a}.

The FSS function $\xi(2L)/\xi(L)$ for the icosahedron and dodecahedron
models \cite{HN-01a} is shown in fig.\ref{fss}. Larger symbols
refer to larger correlation length, the largest one being 
$\xi \approx 300$. 
The O(3) curve is taken from \cite{CEPS-95}.
The comparison shows that the icosahedron and dodecahedron
models approach the O(3) result with increasing correlation length.
(The results for tetrahedron, octahedron and cube -- not shown here --
depart from it.)

\DOUBLEFIGURE[tpb]{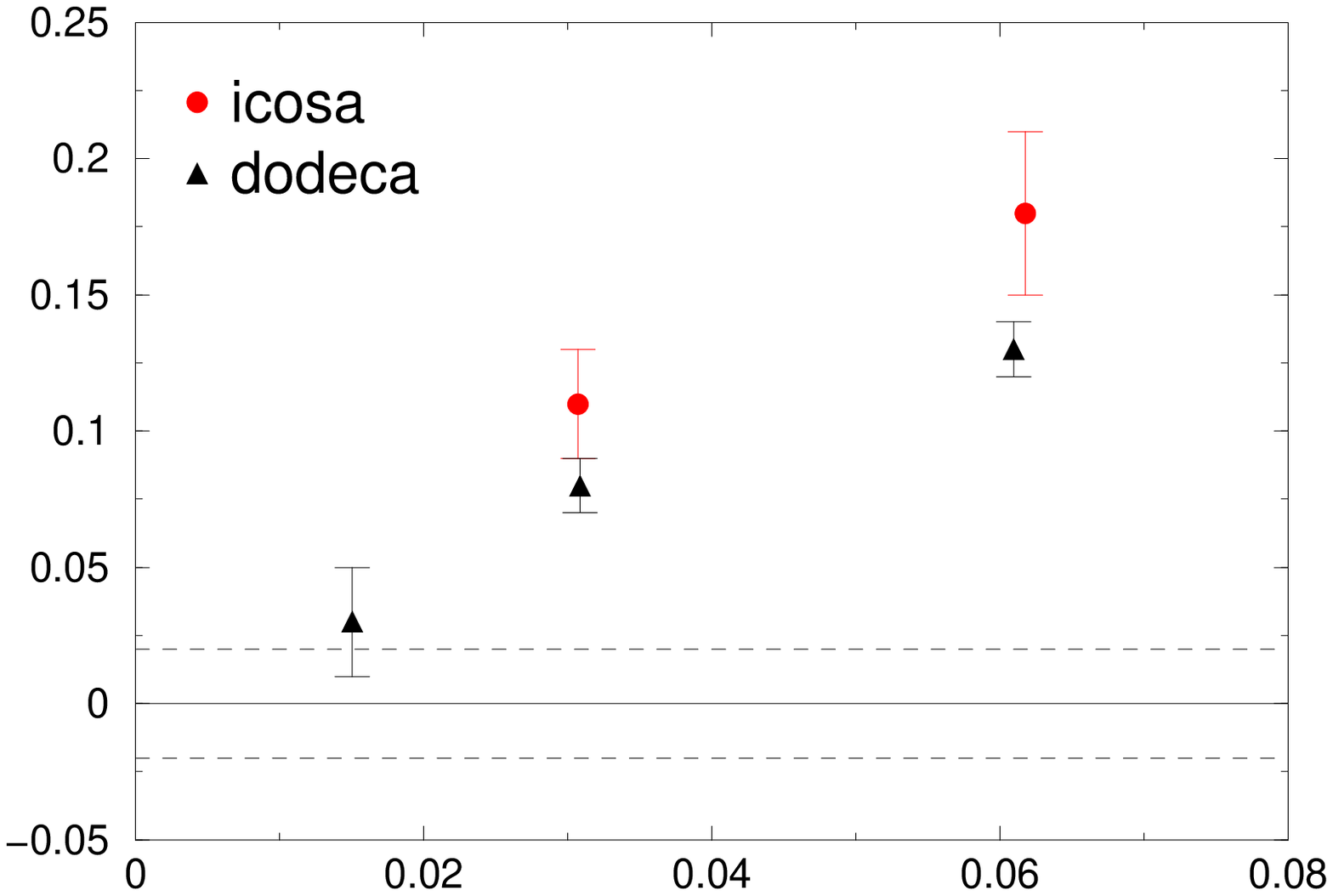,width=1.0\linewidth}%
{fss3b.eps,width=1.0\linewidth}%
{Deviation of $g_R(z)$ from the O(3) result vs. $1/\xi(L)$. 
\label{gr_dev}}%
{The FSS function $\xi(2L)/\xi(L)$ vs $1/z=\xi(L)/L$.
\label{fss}}

The restoration of the O(3) symmetry can be tested, however,
in a more direct way by measuring the direction of average magnetization 
in a {\em fixed physical volume} and checking whether its distribution
becomes uniform in the continuum limit or still prefers the original 
discrete directions.
To do this we introduce the quantity
\be
I_n(\bS) = c \left\{ 
\frac{1}{N}\sum_\bR ( \bR \bS )^n 
-\frac{1}{N'}\sum_{\bR'} ( \bR' \bS )^n \right\} \,,
\ee
where $\bS$ is an arbitrary O(3) vector, $\bR$ runs over
the set of vectors of the actual discrete model, 
$\{ \bR_i\,,\, i=1,\dots,N\}$ 
while $\bR'$ runs over the analogous set of the dual regular solid
(i.e. the unit vectors pointing towards the face centers of the
original regular solid).
For example, in the case of the dodecahedron $N=20$ and the dual 
model is the icosahedron with $N'=12$. 
The value of $n$ is chosen to be the smallest integer 
for which $I_n(\bS)$ is not identically zero. For the dodecahedron
this is $n=6$. Obviously, the integral of $I_n(\bS)$ over
$\bS$ is zero, and $I_n(\bS)$ preserves the discrete 
symmetry ${\cal D}$.
We fix the normalization by the condition $I_n(\bR)=1$
(where $\bR$ is one of the discrete directions).
Up to the normalization constant our $I_n(\bS)$ coincides with
the one introduced by Caracciolo et al. \cite{CMP-01a}.\footnote{Although 
this is not directly seen since they used a concrete representation
for the discrete spin vectors.}

Consider now a fixed physical volume given by $z=L/\xi(L)$
and define the anisotropy
\be
A(z)=\frac{\langle I_6(\bM) \rangle}{\langle \bM^2 \rangle^3}
\,, \qquad {\rm where }\quad \bM=\sum_x \bR(x) \,.
\ee
This quantity is a direct measure of the possible restoration 
of the O(3) symmetry in the continuum limit.
Fig.~3 shows the anisotropy $A(z)$ for the dodecahedron
at $z=1.888,\, 1.668$ and $1.525$ against $1/\xi$.
As the plot shows, $A(z)$ decreases with increasing $\xi$, at least
up to our last value, $\xi \approx 1000$.
This means that the dodecahedron model approaches the O(3) model.
If the decrease persists and $\lim_{\xi\to\infty} A(z;\xi)=0$ then they 
are equivalent in the continuum limit.

Similarly to Caracciolo et al. \cite{CMP-01a} we also considered 
the mixed model
\be
{\cal A}(\bS) = 
-\frac{1}{g} \sum_{x,\mu} \bS(x) \bS(x+\hat{\mu})
-h \sum_x I_n(\bS(x)) \,,
\ee
which interpolates between the O(3) model ($h=0$)
and the discrete model ($h=\infty$).
Here one expects a critical line $g_c(h)$ connecting 
$g_c(0)=0$ with the critical point $g_c(\infty)$ 
of the discrete model. For $g < g_c(h)$ the discrete symmetry
${\cal D}$ is spontaneously broken, while for $g > g_c(h)$
the system is in its symmetric phase.
Since (by the standard wisdom) the correlation length at $g>0$,
$h=0$ is finite, one expects that for $h \ll m^2(g)$
(where $m(g)=1/\xi(g)$ is the inverse of the O(3) correlation
length) no phase transition can occur. This is because the effect
of the corresponding term on the independently fluctuating
regions of size $\xi(g)$ is negligible. (The situation here is in
sharp contrast to the O(2) case where in the massless phase an
arbitrarily small external field can have a drastic effect
on the system in an infinite volume.)

One can study the RG flow in the $(g,h)$ plane. The scenario
advocated by Caracciolo et al. \cite{CMP-01a} means that
the RG flow along the critical line $g_c(h)$ is upwards, towards
the $h=\infty$ point, while if the two models are in the same 
universality class the flow should be directed downwards, towards
$g=h=0$.
A decisive answer would be provided by measuring the flow in the
vicinity of the critical line. For example, consider a point
$(g,h)$ corresponding to a fixed physical volume, $z=L/\xi(L)$
and with anisotropy $A(z)$. To determine the direction of the flow
one has to find a point $(g',h')$ and size $L'$ such that 
$\xi'(L')=s\xi(L)$, $L'=sL$ (i.e. $z'=z$) and $A'(z)=A(z)$,
with a scale factor $s<1$.
At present we have data only for a few points, with $s=1/2$,
at $z\approx 1.9$. The matching pairs $(g,h) \to (g',h')$ are 
shown in fig.~\ref{RG_flow} where the values of $L$ and $L'$
are also indicated. Accordingly, our largest correlation length
was $\xi(L)\approx 60$ which corresponds to $\xi \approx 130$.
(The dashed line representing $g_c(h)$ is drawn in an ad hoc way,
the value $\beta_c(\infty)=1/g_c(\infty)\approx 2.15$ is taken
from \cite{PRS-91}.) 
The pattern in fig.~\ref{RG_flow} suggests a downwards flow,
but one would need points at larger correlation 
length to get a convincing answer.

\DOUBLEFIGURE[tpb]{A_vs_xi.eps,width=1.0\linewidth}%
{RG_flow.eps,width=1.0\linewidth}%
{The anisotropy $A(z)$ vs. $\xi$. \label{A_vs_z}}%
{RG flow in the $(g,h)$ plane. On the vertical axis the values 
of $h/(h+1)$ are plotted.\label{RG_flow}}


Now we would like to discuss the scenario suggested by Caracciolo
et al.\cite{CMP-01a,CMP-01b}.
They expand the correlation functions (defined in a finite $L\times L$
volume) in double power series in $g$ and $h$. Although the
coefficients of this expansion diverge as $L\to\infty$ the functions
appearing in the corresponding RG equation stay finite in this limit.
The assumption they implicitly make is that the RG equation obtained 
remains valid if both $g$ and $h$ are small enough independently.
Below we shall argue that the RG flow obtained by PT is valid
only in a small wedge around the $h=0$ axis, for $h \ll 1/R^2$
where $R=\min\{ \xi(g),L \}$. In the extreme case of $g=0$,
$L={\rm finite}$ (and for simplicity with the usual magnetic field
coupled to $S_z(x)$) one has 
$\langle S_z \rangle = 1/\tanh(hL^2) - 1/(hL^2)$. 
For fixed $L$ this has a power series expansion in $h$ 
with infrared divergent coefficients.
This expansion becomes, however, invalid for $hL^2 \gg 1$ where 
the true answer is $\langle S_z \rangle \approx 1 - 1/(hL^2)$. 
A similar behaviour is expected for finite $g$, with $L$ replaced by 
$\min\{ \xi(g),L \}$.
The reason for the existence of these two regimes is obvious: 
the system behaves differently depending on whether the external 
field is strong enough to keep the (independently fluctuating) regions 
of size $R$ in the preferred direction or not.

The argument given above is also valid for an $h I_n(\bS)$ 
term in the action. There is, however, an important difference:
in the latter case the remnant discrete symmetry is spontaneously
broken at $h > h_c(g)$. This breaking is associated with 
tunneling between different minima of $-h I_n(\bS)$, i.e. the
dodecahedron directions. It is quite obvious that this tunneling
cannot be described by PT: one can replace $-I_n(\bS)$ by a function
which has one single minimum in one of the preferred discrete directions
and has the same perturbative expansion around this minimum as
$-I_n(\bS)$. PT will not notice any difference between the two
cases while in the second case no discrete symmetry is left
hence no phase transition will occur by increasing $h$.
Although these are only speculative arguments, we think they could
be made more rigorous.

As mentioned above, in \cite{CMP-01b} the authors also found 
that the FSS function for the dodecahedron starts to show a deviation 
from the O(3) model at huge correlation length, $\xi \approx 10^5$.
Our criticism is the same as given by Patrascioiu
and Seiler \cite{PS-01}: 1) for $\xi \approx 10^5$
and $L=150$ the physical size of the system is practically zero 
and this small distances are dominated still by cutoff effects, 
2) at such huge estimated $\xi$ one cannot be sure that the system
is in the symmetric phase.

Summarizing, we think that the spontaneous breaking of the discrete 
symmetry is a non-perturbative critical phenomenon, even in the 
vicinity of the $g=h=0$ point.
Concerning the numerical evidence -- in particular the anisotropy 
measurements up to $\xi \approx 1000$ and the fact that in this range
of $\xi$ Caracciolo et al. do not see either a deviation in 
the FSS behaviour --  indicate that the dodecahedron 
(and the icosahedron) models describe the same continuum theory 
as the O(3) model. The question, however, deserves further numerical
study.

The authors thank J\'anos Balog, Sergio Caracciolo, Martin L\"uscher,
Andrea Montanari, Andrea Pelissetto and Peter Weisz for useful
discussions and correspondence. 
This work is supported in part by Schweizerische Nationalfonds 
and by the European Community's Human Potential Programme 
under HPRN-CT-2000-00145 Hadrons/Lattice QCD, BBW Nr. 990143.


\begin{thebibliography}{99}           

\bibitem{PS-98a}
A.~Patrascioiu and E.~Seiler,
\plb{430}{1998}{314}, \heplat{9706011}.

\bibitem{HN-01a}
P.~Hasenfratz and F.~Niedermayer,
\npb{596}{2001}{481}, \heplat{0006021};\\
\npps{94}{2001}{575}, 
\heplat{0011056}.

\bibitem{LWW}
M.~L\"uscher, P.~Weisz and U.~Wolff,
\npb{359}{1991}{221}.


\bibitem{CMP-01a}
S. Caracciolo, A. Montanari, and A. Pelissetto,
\plb{513}{2001}{223}, \heplat{0103017}.

\bibitem{CMP-01b}
S. Caracciolo, A. Montanari, and A. Pelissetto,
This conference, \heplat{0110221}.

\bibitem{CEPS-95}
S.~Caracciolo, R.~G.~Edwards, A.~Pelissetto, and A.~D.~Sokal,
\prl{75}{1995}{1891}, \heplat{9411009};
%
S.~Caracciolo, R.~G.~Edwards, S.~J.~Ferreira, A.~Pelissetto, and A.~D.~Sokal,
\prl{74}{1995}{2969}, \heplat{9409004}.

\bibitem{PRS-91}
A.~Patrascioiu, J.~L.~Richard and E.~Seiler,
\plb{254}{1991}{173}.   

\bibitem{PS-01}
A.~Patrascioiu and E.~Seiler,
\heplat{0110213}.

\end{thebibliography}
\end{document}